\documentclass[12pt,superscriptaddress,floatfix,a4paper]{article}
\usepackage{graphicx,times,amsmath,amsfonts,amssymb,amsthm,bm,bbm,longtable,dcolumn}
\usepackage{revsymb,wasysym,mathrsfs,epsf,color}

\newcommand{\beq}{\begin{equation}}
\newcommand{\eeq}{\end{equation}}
\newcommand{\pt}{\partial}

\begin{document}

\title{\large \bf Reassessing thermodynamic and dynamic constraints on global wind power}

\author{Anastassia Makarieva$^{1,2}$, Victor Gorshkov$^{1,2}$,
Andrei Nefiodov$^{1}$\thanks{Correspondence to:
A.V.~Nefiodov, Theoretical Physics Division, Petersburg Nuclear Physics Institute,
188300 Gatchina, St.~Petersburg, Russia. E-mail: anef@thd.pnpi.spb.ru},\\ Douglas Sheil$^{3}$, Antonio Donato Nobre$^{4}$,
Bai-Lian Li$^{2}$}

\date{\vspace{-5ex}}

\maketitle

\noindent
$^{1}$Theoretical Physics Division, Petersburg Nuclear Physics Institute,
188300 Gatchina, St.~Petersburg, Russia, $^{2}$XIEG-UCR International Center for Arid Land Ecology,
University of California, Riverside 92521-0124, USA,
$^{3}$Norwegian University of Life Sciences, \AA s, Norway,
$^{4}$Centro de Ci\^{e}ncia do Sistema Terrestre INPE,
S\~{a}o Jos\'{e} dos Campos SP 12227-010, Brazil

\begin{abstract}
Starting from basic physical principles, we present a novel derivation linking the global wind power
to measurable atmospheric parameters. The resulting expression distinguishes three components of the atmospheric po\-wer
(the kinetic power associated with horizontal and vertical motion and the gravitational power of precipitation) and highlights
problems with previous approaches. Focusing on Lalibert\'{e} \emph{et al.} (2015), we show how inappropriate treatment of material
derivatives in the presence of phase transitions leads to significant errors in wind power analyses. We discuss the
physical constraints on global wind power and the opportunities provided by considering the dynamic effects of water
vapor condensation.
\end{abstract}

\section{Introduction}

Atmospheric scientists have long sought fundamental principles which can explain global circulation power. This search
has gained renewed significance in view of the need to reconcile an apparent mismatch between model predictions and observed
trends (e.g., Kociuba and Power, 2015).  While models suggest circulation should slow as global temperatures increase,
independent observations indicate that in fact global wind power is increasing
(e.g., de Boiss\'{e}son \emph{et al.,} 2014). Without a coherent theoretical framework such
results cannot be readily reconciled.  Meanwhile no expression for global atmospheric power
in a moist atmosphere has been derived from fundamental physical principles.  Here we derive such an expression.
This permits us to examine the determinants of the atmospheric power budget and to identify and address some widespread misunderstandings.
In particular we argue that equilibrium thermodynamics offer little insight into the constraints on global circulation power.

\section{What is atmospheric power?}

\subsection{Total power}

Work in the atmosphere is done by expanding air. Atmospheric water in its solid and liquid form is incompressible and neither performs work nor
occupies any appreciable volume (Pelkowski and Frisius, 2011).
Work per unit time (power) of an air parcel containing $\tilde{N}$ moles and occupying volume $\tilde{V}$ (m$^3$)
is
\beq
\label{work}
p \frac{d \tilde{V}}{dt} = p \tilde{N} \frac{dV}{dt} + pV \frac{d\tilde{N}}{dt} = \tilde{N} \left(- V\frac{dp}{dt} + R\frac{dT}{dt}\right) +
RT \frac{d\tilde{N}}{dt}.
\eeq
Here we have used the ideal gas law, $pV = RT$ and $pdV = RdT - Vdp$, where $T$ is temperature,
$V \equiv N^{-1}$ is the atmospheric volume occupied by one mole of air, $N$ is air molar density (mol~m$^{-3}$),
$\tilde{V} \equiv \tilde{N} V$, $p$ is air pressure and $R=8.3$ J~mol$^{-1}$~K$^{-1}$ is the universal gas constant.

In Eq.~(\ref{work}) the expression in braces represents part of the parcel's power per mole
that is independent of the rate of phase transitions $d\tilde{N}/dt$. To express it per unit volume
we use the relationship $\tilde{N} \equiv \tilde{V}N$. Then total
power $W_{tot}$  (W) in an atmosphere containing $n$ air parcels with total fixed volume $\mathcal{V} = \sum_{i=1}^n \tilde{V}_i = \int_\mathcal{V}
d\mathcal{V}$ (subscript $i$ refers to the $i$-th air parcel)
is
\beq
\label{workt}
W_{tot} \equiv \sum_{i=1}^n p_i \frac{d \tilde{V}_i}{dt} = \int_\mathcal{V} \left(-\frac{dp}{dt} + NR \frac{dT}{dt} + RT \dot{N}\right) d\mathcal{V}.
\eeq
Here $\dot{N}$ is the molar rate of phase transitions per unit volume (mol~m$^{-3}$~s$^{-1}$). Its integral over volume $\mathcal{V}$ is equal
to the total rate of phase transitions in all the $n$ air parcels: $\int_\mathcal{V} \dot{N}d\mathcal{V} \equiv \sum_{i=1}^n d\tilde{N}_i/dt$.

We will now use the definition of material derivative for $X = \{p, T, N\}$:
\beq
\label{D}
\frac{dX}{dt} \equiv \frac{\pt X}{\pt t} + \mathbf{v}\cdot \nabla X,
\eeq
the continuity equation:
\beq
\label{cont}
\nabla \cdot (N \mathbf{v}) = \dot{N} - \frac{\pt N}{\pt t},
\eeq
and the divergence theorem:
\beq
\label{div}
\int_{\mathcal V} N \mathbf{v} \cdot \nabla T d{\mathcal V}
= \int_{\mathcal{S}} NT (\mathbf{v} \cdot \mathbf{n}) d\mathcal{S} - \int_{\mathcal V} T \nabla \cdot (N\mathbf{v}) d{\mathcal V}=
- \int_{\mathcal V} T \nabla \cdot (N\mathbf{v}) d{\mathcal V}.
\eeq
Here $\mathbf{v}$ is air velocity, $\mathbf{n}$ is unit vector perpendicular to the unit surface of area $d{\mathcal S}$; the integral
is taken over the surface enclosing the atmospheric volume. Since the air does not leave the atmosphere,
we have $\mathbf{v} \cdot \mathbf{n} = 0$, and the integral over $\mathcal{S}$ in (\ref{div}) is zero.
With help of Eqs.~(\ref{D})-(\ref{div}) and the ideal gas law $\pt p/\pt t = R\pt (NT)/\pt t$ we obtain from Eq.~(\ref{workt})
\beq
\label{wtot}
W_{tot} = -\int_\mathcal{V} (\mathbf{v} \cdot \nabla p) d\mathcal{V} = -\int_\mathcal{V} (\mathbf{u} \cdot \nabla p + \mathbf{w}\cdot \nabla p) d\mathcal{V},
\eeq
where $\mathbf{u}$ and $\mathbf{w}$ are the horizontal and vertical air velocities, $\mathbf{v} = \mathbf{u} + \mathbf{w}$.

\subsection{Kinetic power}

In Eq.~(\ref{wtot}) the expression $-(\mathbf{v} \cdot \nabla p)$ represents work performed per unit time per unit air volume
by the pressure gradient. The horizontal pressure gradient generates the kinetic energy of the horizontal wind.
The vertical pressure gradient generates the kinetic energy of the vertical wind plus it changes
the potential energy of air in the gravitational field.

In hydrostatic equilibrium of air we have
\beq\label{he}
\nabla_z p = \rho {\mathbf g},
\eeq
where $\rho = N M$ is air density (kg~m$^{-3}$), $M$ is air molar mass (kg~mol$^{-1}$). In the real atmosphere due to the presence of non-gaseous water the air distribution deviates
from Eq.~(\ref{he}) such that we have $\nabla_z p = (\rho + \rho_l) {\mathbf g}$ and in Eq.~(\ref{wtot})
$-\mathbf{w}\cdot \nabla p =  -\rho \mathbf{w} \cdot \mathbf{g}-\rho_l \mathbf{w} \cdot \mathbf{g}$,
where $\rho_l$ is mass density of the non-gaseous water in the air.

Term $-\rho \mathbf{w} \cdot \mathbf{g}$ represents the vertical flux of air: it is positive (negative) for the ascending (descending)
air. Recalling that $\mathbf{g} = -g\nabla z$ and using
the divergence theorem and the stationary continuity equation, see (\ref{cont}), we can write
\beq
\label{w}
W_P \equiv -\int_{\mathcal{V}} \rho \mathbf{w} \cdot \mathbf{g} d\mathcal{V} = \int_{\mathcal{S}} \mathbf{n} \cdot (\rho \mathbf{w} g z) d\mathcal{S} -
\int_{\mathcal{V}} g z \nabla \cdot (\rho \mathbf{w}) d\mathcal{V} = -\int_{\mathcal{V}} g z \dot{\rho} d\mathcal{V}.
\eeq
For a dry atmosphere where $\dot{\rho} = 0$, this last integral in Eq.~(\ref{w}) is zero and $W_P = 0$: indeed, in this case
at any height $z$ there is as much air going upwards as there is going downwards.
In a moist atmosphere, evaporation $\dot{\rho} > 0$
makes a negative contribution to $W_P$, while condensation $\dot{\rho} < 0$ makes a positive contribution.
This is because $W_P$ reflects the work of water vapor as it travels from
the level where evaporation occurs (where water vapor arises) to the level where condensation occurs (where water vapor disappears).
When condensation occurs above where evaporation occurs, the water vapor expands
as it moves upwards towards condensation, and the work is positive. If condensation occurs below where evaporation occurs, the water vapor must compress
to reach the evaporation site; thus, the work is negative.
When evaporation occurs at the Earth's surface $z = 0$,
$W_P$ (\ref{w}) is equal to $P g H_PS_E$, where $H_P$ is the mean height of condensation, $P$
is global mean precipitation at the surface (kg~m$^{-2}$~s$^{-1}$) and $S_E$ is Earth's surface area
(Makarieva \emph{et al.}, 2013a).
It is natural to call $W_P$ the "gravitational
power of precipitation".

Thus for the stationary total power we have
\begin{eqnarray}
\label{wtotf}
W_{tot} &=& W_K + W_P,\\ \label{WK}
W_K &=& -\int_{\mathcal{V}} (\mathbf{u}\cdot \nabla p + \rho_l \mathbf{w} \cdot \mathbf{g}) d\mathcal{V} \approx
-\int_{\mathcal{V}} \mathbf{u}\cdot \nabla p d\mathcal{V},\\
\label{WP}
W_P &=& -\int_{\mathcal{V}} g z \dot{\rho} d\mathcal{V} \approx P g H_P S_E, \quad  P \equiv -\int_{z > 0} \dot{\rho} d\mathcal{V}/S_E.
\end{eqnarray}
Per unit area, global $W_K$ and $W_P$ were estimated at $2.5$ and $0.8$~W~m$^{-2}$, respectively (Huang and McElroy, 2015; Makarieva \emph{et al.}, 2013a).

The term $-\rho_l \mathbf{w} \cdot \mathbf{g}$ in Eq.~(\ref{WK}) is not related to the gravitational power of precipitation. It describes
kinetic energy generation
on the vertical scale of the order of the atmospheric scale height $H \equiv -p/(\pt p/\pt z) = RT/(Mg) \approx 10$~km.
This energy is generated because the vertical air distribution deviates from the hydrostatic equilibrium (\ref{he}). Hydrometeors act as resistance
not allowing the non-equilibrium pressure difference $\Delta p \sim \rho_l g H$
to be converted to the kinetic energy of the vertical wind.
In the atmosphere on average $\rho_l/\rho \sim 10^{-5}$ (Makarieva \emph{et al.}, 2013a). Without hydrometeors, a pressure difference $\Delta p \sim 10^{-5} \rho g H \sim 1$~hPa would produce
a vertical velocity of about $w \sim 1$~m~s$^{-1}$ ($\rho w^2/2 = \Delta p$). This is two
orders of magnitude larger than the characteristic vertical velocities $w \sim 10^{-2}$~m~s$^{-1}$ of large-scale air motions. Hydrometeors
thus have an effect similar to turbulent friction at the surface which does not allow horizontal velocities to develop. For example, the observed meridional
surface pressure differences of the order of $\Delta p \sim 10$~hPa in the tropics, if friction were absent, would have produced horizontal air velocities
of about 40~m~s$^{-1}$ instead of the observed $7$~m~s$^{-1}$.
The term $-\rho_l \mathbf{w} \cdot \mathbf{g}$
is less than 1\% of $W_{tot}$ and can be neglected: its volume integral taken per unit surface area is
less than $\rho_l g H w \sim 10^{-5} p w \sim 10^{-2}$~W~m$^{-2}$, where $p = \rho g H = 10^5$~Pa is air pressure at the surface.

\section{Revisiting the current understanding of the atmospheric power budget}

\subsection{The physical meaning of $W_{tot}$, $W_K$ and $W_P$}

To our knowledge, Gorshkov (1982) was the first to estimate $W_P$ for land assuming $H_P = 2$~km.
In the meteorological literature,
Pauluis \emph{et al.} (2000) defined precipitation-related frictional dissipation as
$W_P + \int_V \rho_l \mathbf{w} \cdot \mathbf{g} d\mathcal{V}$ and estimated its value for the tropics. This estimate was later revised
by Makarieva \emph{et al.} (2013a). Pauluis \emph{et al.} (2000) stated that in the stationary case total power of
atmospheric motions must be equal to the sum of the dissipation rate at the microscale around the hydrometeors
and the rate of dissipation of motions on convective scale. The latter was approximated it as the power of
the buoyancy force that is proportional to vertical velocity (Pauluis \emph{et al.}, 2000, their Eq.~8).
Derivation of Eqs.~(\ref{wtotf})-(\ref{WP}) have not been previously published.

These equations provide insight into the controls on circulation power.
First, we can see that $W_P$ does not depend on the interaction between the air and the falling drops. This term would be present in the atmospheric
power budget even if drops were experiencing free fall
and did not interact with the air at all (such that no frictional dissipation on drops occurred in the atmosphere). Second, the small term $-\rho_l \mathbf{w} \cdot \mathbf{g}$ describes
generation of kinetic energy at the convective scale $H$ (while this energy dissipates on hydrometeors) and thus should be included into
the total rate of kinetic energy generation $W_K$.
Finally, in the absence of a derivation relating Eq.~(\ref{work}) and Eq.~(\ref{wtotf}), it is not clear {\it a priori} whether
precipitation makes an independent contribution to the total power budget. Indeed it could appear as a form of turbulent dissipation
of the kinetic energy generated at larger scales (i.e. water is lifted by turbulent diffusion),
and thus be comprised in the expression for the total kinetic power.

We disagree with the recent commentary of Pauluis (2015) on the work of Lalibert\'{e} \emph{et al.} (2015).
To provide some context, an ideal atmospheric Carnot cycle consuming heat flux $F = 100$~W~m$^{-2}$ ($F$ is
limited from above by solar power flux reaching the planetary surface) at surface temperature $T_{in} = 300$~K and releasing
heat at $T_{out} = T_{in}- \Delta T $ with $\Delta T = 30$~K being roughly the magnitude of the Earth's
greenhouse effect, would generate kinetic energy at a rate of $W_C = F (\Delta T/T_{in}) = 10$ W~m$^{-2}$.
Lalibert\'{e} \emph{et al.} (2015) estimated total atmospheric power $W_{tot}$ per unit area to be around 4~W~m$^{-2}$.
Comparing this result with $W_C$, Pauluis (2015) notes that "estimates for the rate of kinetic
energy production by atmospheric motions are about half this figure". Confusion has apparently arisen
between total atmospheric power $W_{tot}$ and kinetic power $W_K$. Indeed, Pauluis (2015) continues
that "the difference is very likely due to Earth's hydrological cycle, which reduces the
production of kinetic energy in two ways", one of which is the gravitational power of precipitation $W_P$.
However, as is clear from Eq.~(\ref{wtotf}), $W_P$ is equal to the difference between $W_{tot}$ and $W_K$
and is unrelated to the difference $W_C - W_{tot}$.

Since the relationship between $W_C$, $W_P$ and $W_K$ was discussed by Pauluis \emph{et al.} (2000) (see their Eq.~(7)),
the confusion may stem from a misinterpretation of the magnitude estimated by Lalibert\'{e} \emph{et al.} (2015)
as atmospheric power, $\int_\mathcal{M} (1/\rho) (dp/dt) d\mathcal{M} = \int_\mathcal{V} (dp/dt) d\mathcal{V}$, for $W_K$ --
while it was in fact $W_{tot}$. Notably, neither
Pauluis \emph{et al.} (2000) nor Lalibert\'{e} \emph{et al.} (2015) identified the fact that
$W_K$ depends on horizontal and not vertical velocities. This is essential for comparing theory and observations.
Horizontal pressure gradients and wind velocities are observed, while the vertical velocities are
inferred with significant uncertainty.

There are other studies where kinetic energy generation is estimated from horizontal velocities
as $W_K$ (\ref{WK}) (see, e.g., Boville and Bretherton, 2003; Huang and McElroy, 2015). At the same time, $W_K$
is sometimes confused for the total atmospheric power: i.e. in the total power budget the gravitational
power of precipitation, $W_P$, is overlooked (e.g., Huang and McElroy, 2015, their Fig.~10).

\subsection{The analysis of Lalibert\'{e} \emph{et al.} (2015)}

Equations~(\ref{work})-(\ref{WP}) also allow us to identify errors in analyses of $W_{tot}$.
To estimate the atmospheric power budget, Lalibert\'{e} \emph{et al.} (2015) adopted the thermodynamic identity
\begin{equation}
\label{fle}
T \frac{ds}{dt} \equiv \frac{dh}{dt} - \frac{1}{\rho} \frac{dp}{dt} + \mu \frac{dq_T}{dt},
\end{equation}
where $s$ is entropy, $h$ is enthalpy, $\mu$ is chemical potential (all per unit mass of wet air),
$1/\rho$ is specific air volume and $q_T$ is water mixing ratio\footnote{The unconventional sign at the chemical potential term follows from $\mu$
being defined in Eq.~(\ref{fle})
relative to dry air: hence, when the dry air content diminishes (while $dq_T/dt>0$) this term is negative. For details see p.~8
in the Supplementary Material of Lalibert\'{e} \emph{et al.} (2015).}.
When integrating Eq.~(\ref{fle}) over atmospheric mass, Lalibert\'{e} \emph{et al.} (2015)
noted that the enthalpy term vanishes, $\int_{\mathcal{M}} (dh/dt) d\mathcal{M} = 0$,
because the atmosphere is in a steady state. However, Eqs.~(\ref{D}) and (\ref{cont}) indicate that for any
scalar quantity, in our case enthalpy $h$, in the stationary case, $\pt h/\pt t = 0$, we have
\beq
\label{X}
\int_{\mathcal M} \frac{dh}{dt} d\mathcal{M} = \int_{\mathcal V} \frac{dh}{dt} \rho d\mathcal{V} = -\int_\mathcal{V} h \dot{\rho} d\mathcal{V}.
\eeq
Here $\dot{\rho}$ is the mass rate of phase transitions per unit volume (kg~m$^{-3}$~s$^{-1}$),
$\nabla \cdot (\rho \mathbf{v}) = \dot{\rho}$. This integral is not zero in an atmosphere
where phase transitions take place.

This omission creates a significant error.
To obtain an approximate estimate we can assume that all evaporation occurs at the surface, while
all condensation occurs at a global mean height of $H_P = 2.5$~km, where about half of surface
water vapor has condensed in the ascending air (see Makarieva \emph{et al.}, 2013a). Then we have
$[\int_{\mathcal{M}} (dh/dt) d\mathcal{M}]/S_E \approx -P [h(0) - h(H_P)]$, see Eq.~(\ref{WP}), where
$h(0)$ and $h(H_P)$ are mean enthalpies at the surface ($z = 0$) and at mean condensation height ($z =H_P$).
With $h \approx c_p T + L q$, where $q$ is water vapor mass mixing ratio, $c_p = 10^3$~J~kg$^{-1}$~K$^{-1}$ is heat capacity of air at constant pressure,
$L = 2.5\times 10^6$~J~kg$^{-1}$ is latent heat of vaporization, $T(0) - T(H_P) \approx 15$~K, $q(0) - q(H_P) = 0.5 q(0) \approx 0.5\times 10^{-2}$,
$P = 10^3$~kg~m$^{-2}$~yr$^{-1}$, we find that $[\int_{\mathcal{M}} (dh/dt) d\mathcal{M}]/S_E \sim -1$~W~m$^{-2}$.
This figure is about one quarter of the total atmospheric power $W_{tot}/S_E \approx 4$~W~m$^{-2}$ estimated by Lalibert\'{e} \emph{et al.} (2015) for
the MERRA re-analysis and the CESM model.

As Eq.~(\ref{fle}) is an identity, whether the incorrect assumption has increased or decreased the value of atmospheric
power depends on the particular method of calculating it.
Lalibert\'{e} \emph{et al.} (2015) first calculated the mass integral of $Tds/dt$ from Eq.~(\ref{fle}),
then calculated $\mu dq_T/dt$ from atmospheric parameters and then used the obtained values to estimate the total power
$-\int_\mathcal{M}(1/\rho)(dp/dt)d\mathcal{M}$. In such a procedure,
putting $\int_{\mathcal{M}} (dh/dt) d\mathcal{M} = 0$ resulted
in an overestimate of the atmospheric power by about $1$~W~m$^{-2}$.
Since the omitted term depends on precipitation rate, its omission is crucial not only
for a correct estimate of the mean value of $W_{tot}$, but also for the determination of any trends related
to precipitation, which is a focus of Lalibert\'{e} \emph{et al.} (2015).
Thus the quantitative conclusions of Lalibert\'{e} \emph{et al.} (2015) appear invalid.

\section{Dynamic versus thermodynamic approach}

Under an assumption of local thermodynamic equilibrium Eq.~(\ref{fle}) defines $ds/dt$ via measurable atmospheric variables.
As such, it does not carry any additional information about the atmosphere
besides that contained in the set of local values of pressure, temperature, relative humidity and air velocity. Total atmospheric power
can be estimated from Eqs.~(\ref{wtotf})-(\ref{WP}) without involving entropy. This fact illustrates a limitation
of the thermodynamic approach in that it cannot explain
why a particular thermodynamic cycle exists: i.e. why does the atmosphere generate power.

Indeed, the kinetic power $W_K$ can be viewed as a measure of the dynamic disequilibrium of the Earth's atmosphere. In equilibrium,
for example, under conditions of hydrostatic and geostrophic or cyclostrophic balance,
no power is generated and $W_K = 0$.
Considerations of entropy allow one to constrain the maximum
work that can be extracted from a given thermodynamic cycle. However, equilibrium thermodynamics cannot
predict whether the considered system will be in equilibrium. It thus provides no insight as to if or why $W_K$ on Earth differs from zero.
The obvious upper limit --
the global efficiency of solar energy conversion into useful work, which amounts to about 90\% (Wu and Liu, 2010; see Pelkowski, 2012 for a rigorous theoretical
discussion) -- appears irrelevant for constraining $W_K$: $W_K$ is far below this thermodynamic limit, being about 1\% of incoming solar
radiation.

In contrast, the dynamic approach identifies situations when production of kinetic energy cannot be zero in principle.
Saturated water vapor in the gravitational field above the liquid ocean is a unique dynamic system, because its
pressure is controlled by temperature alone rather than by temperature and molar density as
for the non-condensable ideal gases. Specifically, while dry air can rise in hydrostatic equilibrium,
the saturated water vapor cannot.
In an atmosphere composed of pure saturated water vapor
the water vapor rises (returning to the Earth as precipitation) and kinetic energy is produced at a local
rate $-w (\pt p_v/\pt z - \rho_v g) > 0$, where $\rho_v$ is mass density of water vapor (Makarieva \emph{et al.}, 2015b).
If non-condensable gases are added to such an atmosphere, it becomes possible to arrange a hydrostatic equilibrium
in the vertical plane, such that all the kinetic power that derives from condensation is now generated in the horizontal plane:
$- \mathbf{u}\cdot \nabla p = -w (\pt p_v/\pt z - p_v/H) = -w R T N \pt \gamma/\pt z$. Since $-\int_{w>0} w N (\pt \gamma/\pt  z) dz \approx P/M_v$,
where $M_v$ is molar mass of water vapor, the global kinetic power that can be derived from condensation is about 4~W~m$^{-2}$
(assuming mean temperature of condensation $T \sim 270$~K and $P = 10^3$~kg~m$^{-2}$~yr$^{-1}$) (Makarieva \emph{et al.}, 2013a).
This theoretical estimate is about 40\% higher than the most recent estimate of $W_K /S_E= 2.5$~W~m$^{-2}$
obtained using the MERRA re-analysis (Huang and McElroy, 2015).

The difference is likely related to an insufficient spatial and temporal resolution.
Even when in real eddies there is no evaporation in the descending air
(i.e. for $w <0$ we have $\pt \gamma/\pt z = 0$ and $\dot{N} = 0$), since in the rising saturated air we always
have $\pt \gamma/\pt z < 0$, the spatial and temporal averaging produces a non-zero vertical gradient
$\pt \gamma/\pt z < 0$ everywhere. This results in "spurious" evaporation $\dot{N} \approx w \pt \gamma/\pt z > 0$ for $z > 0$.
It diminishes the integral of $-\int_{z>0}\dot{N}d\mathcal{V}$ and all the related quantities, including
the kinetic power derived from condensation\footnote{This spurious evaporation also diminishes the integral of $dh/dt$ (\ref{X}),
a problem not discussed by Lalibert\'{e} et al. (2015).}.
Our theoretical estimate $W_K/S_E = 4$~W~m$^{-2}$ suggests that with increasing spatial and temporal resolution
the atmospheric power estimated from observations should grow. In agreement with these ideas,
Kim and Kim (2013) using the daily averaged MERRA data obtained $W_K /S_E= 2$~W~m$^{-2}$, which is
20\% less than the value obtained by Huang and McElroy (2015) using 3-hour observations (Huang J, personal communication).

In global circulation models a non-zero rate of kinetic energy generation is achieved by
introducing an {\it ad hoc} intensity of turbulent diffusion which determines the rate at which kinetic energy is dissipated (and, in the steady state,
generated) (see Makarieva \emph{et al.}, 2015a). It is this parameterization that postulates a certain value of $W_K$ and
controls its behavior. Since this parameterization is unrelated to the hydrological cycle (i.e. one and the same
turbulent diffusion coefficient can be used in both dry and moist models), comparing $W_K$ across models
with varying intensity of the hydrological cycle does not shed light on the actual role of water vapor.
Meanwhile the condensation-induced atmospheric dynamics outlined above (for further details see
Makarieva \emph{et al.}, 2013b; 2015b and references therein)
suggests that in the absence of the hydrological cycle
the atmospheric power on Earth would have been negligible compared to what it is today.
We thus urge attention to the dynamic effects of condensation.

\section*{Acknowledgements}

We thank Dr.~Olivier Pauluis, Dr.~Fr\'{e}d\'{e}ric Lalibert\'{e} and five anonymous reviewers for their critical comments.
This work is partially supported by Russian Scientific Foundation under Grant no.~14-22-00281 and
the University of California Agricultural Experiment Station.

\noindent
{\bf  References}

\noindent
Boville BA, Bretherton CS. 2003. Heating and kinetic energy dissipation in the NCAR Community Atmosphere Model. \emph{Journal of Climate} {\bf  16}: 3877--3887.

\noindent
de Boiss\'{e}son E, Balmaseda MA, Abdalla S, K\"{a}ll\'{e}n E, Janssen PAEM. 2014.
How robust is the recent strengthening of the Tropical Pacific trade winds?
\emph{Geophysical Research Letters} {\bf  41}: 4398--4405.

\noindent
Gorshkov V.G. 1982: Energetics of the biosphere. Leningrad Politechnical Institute, 80 pp.

\noindent
Huang J, McElroy MB. 2015. A 32-year perspective on the origin of wind energy in a warming climate.
\emph{Renewable Energy} {\bf  77}: 482--492.

\noindent
Kim YH, Kim MK. 2013. Examination of the global lorenz energy cycle using MERRA and NCEP-reanalysis 2.
\emph{Climate Dynamics} {\bf  40}: 1499--1513.

\noindent
Kociuba G, Power SB. 2015. Inability of CMIP5 models to simulate recent strengthening of the Walker circulation: implications for projections. \emph{Journal of Climate} {\bf  28}: 20--35.

\noindent
Lalibert\'{e} F, Zika J, Mudryk L, Kushner PJ, Kjellsson J, D\"{o}\"{o}s K. 2015.
Constrained work output of the moist atmospheric heat engine in a warming climate.
\emph{Science} {\bf  347}: 540--543.

\noindent
Makarieva AM, Gorshkov VG, Nefiodov AV, Sheil D, Nobre AD, Bunyard P, Li BL. 2013a.
The key physical parameters governing frictional dissipation in a precipitating atmosphere.
\emph{Journal of the Atmospheric Sciences} {\bf  70}: 2916--2929.

\noindent
Makarieva AM, Gorshkov VG, Sheil D, Nobre AD, Li BL. 2013b.
Where do winds come from? A new theory on how water vapor condensation influences atmospheric pressure and dynamics.
\emph{Atmospheric Chemistry and Physics} {\bf  13}: 1039--1056.

\noindent
Makarieva AM, Gorshkov VG, Nefiodov AV, Sheil D, Nobre AD, Li BL. 2015a.
Comment on "The tropospheric land-sea warming contrast as the driver of tropical sea level pressure
changes". \emph{Journal of Climate} {\bf  28}: 4293--4307, doi:10.1175/JCLI-D-14-00592.1.

\noindent
Makarieva AM, Gorshkov VG, Nefiodov AV. 2015b.
Empirical evidence for the condensational theory of hurricanes. \emph{Physics Letters A},  {\bf  379}: 2396--2398,   doi:10.1016/j.physleta.2015.07.042.

\noindent
Pauluis O, Balaji V, Held IM. 2000. Frictional dissipation in a precipitating atmosphere.
\emph{Journal of the Atmospheric Sciences} {\bf  57}: 989--998.

\noindent
Pauluis OM. 2015. Atmospheric science. The global engine that could. \emph{Science}  {\bf 347}: 475--476. 
doi:10.1126/science.aaa3681.

\noindent
Pelkowski J, Frisius T. 2011. The theoretician's clouds -- heavier or lighter than air? On densities in atmospheric thermodynamics. \emph{Journal of the Atmospheric Sciences} {\bf  68}: 2430--2437.

\noindent
Pelkowski J. 2012. Of entropy production by radiative processes in a conceptual climate model.
\emph{Meteorologische Zeitschrift} {\bf  21}: 439--457.

\noindent
Wu W, Liu Y. 2010. Radiation entropy flux and entropy production of the Earth system.
\emph{Reviews of Geophysics} {\bf  48}: RG2003.

\end{document}